\documentclass[12pt,a4paper]{article}

\usepackage{amsmath,amssymb,epsfig,cite}

\def\lp{\left(}
\def\rp{\right)}
\def\lb{\left[}
\def\rb{\right]}
\def\be{\begin{equation}}
\def\ee{\end{equation}}
\begin{document}

\title{Charged shells in a $(2+1)$-dimensional spacetime} 
\author{Ernesto F. Eiroa$^{1,2,}$\thanks{e-mail: eiroa@iafe.uba.ar}, 
Claudio Simeone$^{2,3,}$\thanks{e-mail: csimeone@df.uba.ar}\\
{\small $^1$ Instituto de Astronom\'{\i}a y F\'{\i}sica del Espacio,}\\
{\small Casilla de Correo 67, Sucursal 28, 1428, Buenos Aires, Argentina}\\
{\small $^2$ Departamento de F\'{\i}sica, Facultad de Ciencias Exactas y 
Naturales,} \\ 
{\small Universidad de Buenos Aires, Ciudad Universitaria Pabell\'on I, 1428, 
Buenos Aires, Argentina} \\
{\small $^3$ IFIBA, CONICET, Ciudad Universitaria Pabell\'on I, 1428, 
Buenos Aires, Argentina}} 
\maketitle

\begin{abstract}

We study circular shells in a $(2+1)$-dimensional background within the framework of Einstein--Born--Infeld theory. For shells around black holes we analyze the  mechanical stability under perturbations preserving the symmetry. Shells around vacuum are also discussed. We find a large range in the values of the parameters compatible with stable configurations.  \\

\noindent PACS number(s): 04.20.Gz; 04.40.-b; 04.70.Bw 

\end{abstract}

\section{Introduction}

The Born--Infeld electromagnetic theory \cite{borninf} was introduced in order to avoid the infinite self energy of a charged point particle, which constitutes a well known problem within Maxwell electrodynamics, and it is the only non--linear theory without birefringence. Maxwell and Born--Infeld theories have electric--magnetic duality invariance \cite{gibbons}, a property not shared with other electromagnetic theories. Recently, the Born--Infeld electrodynamics has received considerable attention as it can be obtained as a low energy limit of string theory, which at present constitutes the main candidate for a unified theory. The spherically symmetric solution of Einstein gravity coupled to Born--Infeld electrodynamics corresponds to a charged black hole \cite{hoffmann}. The  solution in $2+1$ dimensions \cite{cg} also represents a black hole which is singular at the origin, where the Ricci scalar and the Ricci square diverge. However, the metric is regular everywhere, and is given by
\be
ds^2=-f(r)dt^2+f^{-1}(r)dr^2+r^2d\varphi^2,\label{1}
\ee
where the function $f(r)$ has the form
\be
f(r)= -M+\frac{r^2}{l^2}+\frac{2r}{b^2} \lp r-\sqrt{r^2+b^2Q^2}\rp-2Q^2\ln\lp\frac{r+\sqrt{r^2+b^2Q^2}}{r_1}\rp.\label{1f}
\ee
The dimensionless constants $M$ and $Q$ are  identified as the ADM mass and the charge, $-l^{-2}=\Lambda$ is the cosmological constant with dimensions $[\rm{length}]^{-2}$, and $b$ is the Born--Infeld parameter; $b$ has dimensions of length, and therefore those of the inverse of the field strength. The limit $b\to 0$ corresponds to Maxwell linear electrodynamics; in this limit the well known static BTZ solution with charge \cite{btz} is recovered, and we simply have $f(r)= -M+ r^2/l^2-2Q^2\ln (r/r_0)$. The spacetime geometry of the BTZ black hole has constant negative curvature and it is, locally, that of anti--de Sitter (AdS) space; it only differs from anti--de Sitter in its global properties. This feature makes the BTZ solution of great interest within the framework of the AdS/CFT correspondence. In the geometry defined by Eqs. (\ref{1}) and (\ref{1f}), the radii $r_0$ and $r_1$ are related by $2\sqrt{e}r_1=r_0$, and are associated to the zero of the electrostatic potential $A^0$ in each theory. We 
will assume that the cosmological constant $\Lambda$ is negative (i.e. $l^2>0$), so that in the case of a vanishing charge the geometry is asymptotically anti-DeSitter. For suitable values of the other parameters, this choice makes possible a standard horizon structure \cite{cg}, in which the metric function is always positive beyond a certain radius (different for each theory). 

Thin layers of matter naturally appear in the context of general relativity and cosmology. For instance, they can be used to model the gravitational collapse to a black hole, and also for the study of the evolution of bubbles and domain walls in a cosmological setting. Spherically symmetric shells around vacuum (bubbles), around black holes and stars, and also being the throat of traversable wormholes, have been extensively studied \cite{sta,stars,wh4}; these shells have also been considered in more than four dimensions \cite{wh5,eisi12a}. The dynamics of collapsing shells in three spacetime dimensions has been presented and applied to several examples in Refs. \cite{crisol}. Besides, shells in a three dimensional background within Einstein--Maxwell theory have been associated to thin-shell wormholes \cite{raha}.  On the other hand, we have recently analyzed the problem in the four dimensional case for Einstein gravity coupled to Born--Infeld electrodynamics in relation with thin-shell wormholes \cite{risi} 
and shells around vacuum or around black holes \cite{eisi-bi}.

The current interest in $(2+1)$-dimensional Einstein--Maxwell and Einstein--Born--Infeld black holes, leads to address the related problem of the behavior of thin layers associated with them. In the present article, we study the characterization and the linearized stability under perturbations preserving the symmetry of circular charged shells around black holes; as a particular case, we also analyze shells around vacuum. We mathematically build the shells starting from two black hole geometries with a negative cosmological constant, given by Eqs. (\ref{1}) and (\ref{1f}). We apply the well known cut and paste procedure; we work under the condition that the resulting shell is constituted by normal matter, that is, by matter satisfying the energy conditions. We set the units so that $G=c=1$.

\section{Mathematical construction}

To define a shell (more precisely a ring) of radius $a$ we take two geometries of the form (\ref{1}) with different metric functions $f_1,\,f_2$ and different coordinates $r_1,\, r_2,\, t_1$ and $t_2$, and we remove the outer region $r_1> a$ of one of them and the inner region $r_2<a$ of the other one. Then, we join the resulting manifolds at $r_1=r_2=a$ to form a new one. For a shell surrounding a black hole, the radius $a$ is chosen beyond the largest horizon radius of the inner geometry and also $a$ should be large enough to remove the horizons (if any) of the outer geometry. In the case of a shell around vacuum, the first restriction is no longer needed. The new complete spacetime in general includes a $1+1$ dimensional matter shell at $r_{1,2}=a$; this implies that the metric of the complete geometry, though continuous everywhere, must have discontinuous derivatives. More precisely, at the shell radius $r_1=r_2=a$ the line element satisfies $f_1(a){dt_1}^2=f_2(a){dt_2}^2$, while the derivatives of the 
metric at both sides of the $1+1$ dimensional surface $r_{1,2}=a$ are related with the surface energy-momentum tensor $S_i^j$  by the Lanczos equations \cite{daris,mus}
\be
8\pi S_i^j=-[K_i^j]+\delta_i^j[K]
\ee
where $K_i^j$ is the extrinsic curvature tensor, $K$ stands for its trace, and the brackets denote the jump of  a given quantity across the circumference $r_{1,2}=a$. We let the radius $a$ to be a function of the proper time $\tau$ on the ring. Then the  components of the extrinsic curvature tensor are given by
\be
{K_\tau^\tau}^{1,2}=-\frac{\ddot a+{f'_{1,2}}(a)/2}{\sqrt{f_{1,2}(a)+{\dot a}^2}},
\ee
\be
{K_\varphi^\varphi}^{1,2}=\frac{1}{a}\sqrt{f_{1,2}(a)+{\dot a}^2},
\ee
where a prime denotes a derivative with respect to $r$ and a dot stands for $d/d\tau$. The surface energy density $\lambda=-S_\tau^\tau$ and the pressure $p=S_\varphi^\varphi$ are given by
\be
\lambda=-\frac{1}{8\pi a}\sqrt{f_2(a)+{\dot a}^2}+\frac{1}{8\pi a}\sqrt{f_1(a)+{\dot a}^2},
\ee
\be
 p=\frac{\ddot a+{f_2}'(a)/2}{8\pi \sqrt{f_2(a)+{\dot a}^2}}-\frac{\ddot a+{f_1}'(a)/2}{8\pi \sqrt{f_1(a)+{\dot a}^2}}.
\ee
The energy and pressure for a static ($a=a_0$) ring read
\be
\lambda_0=-\frac{1}{8\pi a_0}\sqrt{f_2(a_0)}+\frac{1}{8\pi a_0}\sqrt{f_1(a_0)},
\ee
\be
 p_0=\frac{{f_2}'(a_0)}{16\pi \sqrt{f_2(a_0)}}-\frac{{f_1}'(a_0)}{16\pi \sqrt{f_1(a_)}}.
\ee
A static shell of normal (i.e. non exotic) matter satisfies the weak energy condition, i.e. $\lambda_0\geq 0$ and $\lambda_0+p_0\geq 0$; this implies $f_1(a_0)\geq f_2(a_0)$. With some changes, the cut and paste procedure can be used to construct a wormhole geometry; in this case, both terms in the expressions for $\lambda$ and $p$ will have the same sign. Then, the weak energy condition cannot be fulfilled because the joining of two exterior regions $(r_{1,2}>a_0)$ with metrics given by $f_1(r_1)$ and $f_2(r_2)$ would give $\lambda_0=-(\sqrt{f_1(a_0)}+\sqrt{f_2(a_0)})/(8\pi a_0)<0$. In our study of bubbles or shells around black holes, we will work under the assumption of normal matter.

\section{Stability analysis}

From the Lanczos equations, we can obtain an equation for the evolution of the shell radius in the form analogous to the energy conservation for a point particle restricted to a motion in only one spatial dimension. Squaring twice the expression for the energy density and after some algebra we obtain
\be
\dot{a}^{2}+V(a)=0\label{mal},
\ee	
where the potential $V(a)$ has the form 
\be
V(a)=\frac{f_1(a)+f_2(a)}{2}-\lb\frac{f_1(a)-f_2(a)}{16\pi a\lambda (a)}\rb ^{2}-\lb 4\pi a\lambda(a) \rb^{2}.\label{pot}
\ee
The energy and pressure satisfy the conservation equation
\be
\frac{d}{d\tau}(a\lambda)+p\frac{da}{d\tau}=0,
\ee
which leads to
\be
\lambda'=-\frac{1}{a}(\lambda+p).
\ee
If an equation of state $p=p(\lambda)$ is given, this can be integrated to obtain the energy density as a function of the ring radius by inverting the resulting relation
\be
\ln a=-\int\frac{d\lambda}{\lambda+p(\lambda)}+C.\label{sol1}
\ee
Then $\lambda(a)$ can be substituted in $V(a)$ and this would allow, in principle, to obtain the time evolution of the ring radius.

We are interested in a perturbative treatment of the dynamics, more precisely, the  stability of static solutions. Because of the analogy with the problem of a point particle in  a one dimensional potential, the analysis is straightforwardly carried out in terms of the sign of the second derivative of the potential $V(a)$ at an equilibrium configuration given by $a=a_0$. We introduce the definitions
\be
S(a)=\frac{f_1(a)+f_2(a)}{2},
\ee
\be 
R(a)=\frac{f_1(a)-f_2(a)}{2},
\ee
\be
m(a)=2\pi a\lambda .
\ee
This allows to write the potential  as
\be
V(a)=S-\frac{R^2}{16m^2}-4 m^2.
\ee
A static configuration $a=a_0$ implies $V(a_0)=0$, equilibrium requires $V'(a_0)=0$, and the condition for stability  is that $V''(a_0)>0$.
The first and second derivatives of the potential read
\be
V'(a_0)=S'-\frac{R'R}{8}\lp\frac{1}{m}\rp^2-\frac{R^2}{8m}\lp\frac{1}{m}\rp'-8mm',
\ee
\begin{eqnarray}
V''(a_0)&=&S''-8{m'}^2-8mm''-\frac{R^2}{8}\lb{\lp \frac {1}{m}\rp'}^2+\frac{1}{m}{\lp\frac{1}{m}\rp''}\rb\nonumber\\
& &-\frac{1}{4}R'R\frac{1}{m}\lp\frac{1}{m}\rp'-\frac{1}{8}\lp\frac{1}{m}\rp^2\lb {R'}^2+RR''\rb.
\end{eqnarray}
In the perturbative treatment, only the first derivative of the pressure at the equilibrium position is involved in the dynamics. Second derivatives of $m$ and of $m^{-1}$ can be expressed in terms of first derivatives by recalling the conservation equation $a\lambda '= -(\lambda +p)$ and introducing the parameter
\be
\eta\equiv \frac{p'(a_0)}{\lambda'(a_0)}
\ee
which relates the derivatives of the pressure and energy density at the equilibrium radius $a_0$;  in the case $0\leq \eta <1$ this parameter can be  understood as the speed of sound along the ring. With this definition we have
\be
m''=-\frac{1}{a_0}\lp m'-\frac{m}{a_0}\rp\eta,
\ee
\be
\lp\frac{1}{m}\rp''=\frac{{2m'}^2}{m^2}+\frac{1}{m^2a_0}\lp m'-\frac{m}{a_0}\rp\eta.
\ee
 Then taking into account that $V'(a_0)=0$  and defining the auxiliary functions
\be
X(a_0)=\frac{1}{8m}\lb S'-\frac{RR'}{8m^2}-\frac{R^2}{8m}\lp\frac{1}{m}\rp'\rb,
\ee
\be
Y(a_0)=S''-\frac{1}{8m^2}\lp {R'}^2+RR''\rp-\frac{RR'}{2m}\lp\frac{1}{m}\rp'-\frac{R^2}{8}\lp\frac{1}{m}\rp'^2,
\ee
\be
Z(a_0)=\frac{8m}{a_0}\lp\frac{m}{a_0}-m'\rp\eta+\frac{R^2}{8m}\lb\frac{2{m'}^2}{m^3}+\frac{1}{m^2a_0}\lp m'-\frac{m}{a_0}\rp\eta\rb,
\ee
the stability condition for a static equilibrium configuration takes the concise form
\be
Z(a_0)<Y(a_0)-8X^2(a_0).\label{sta}
\ee
Because we are interested in the study of bubbles and rings around black holes (not in thin-shell wormholes), the additional condition of normal matter  must be imposed. The natural way to understand the results is then to present the stability regions by drawing the intersection of the relation (\ref{sta}) and the inequalities $\lambda _0 \geq 0,\ \lambda _0 + p_0 \geq 0$ for different values of the parameters. 

\begin{figure}[t!]
\centering
\includegraphics[width=16cm]{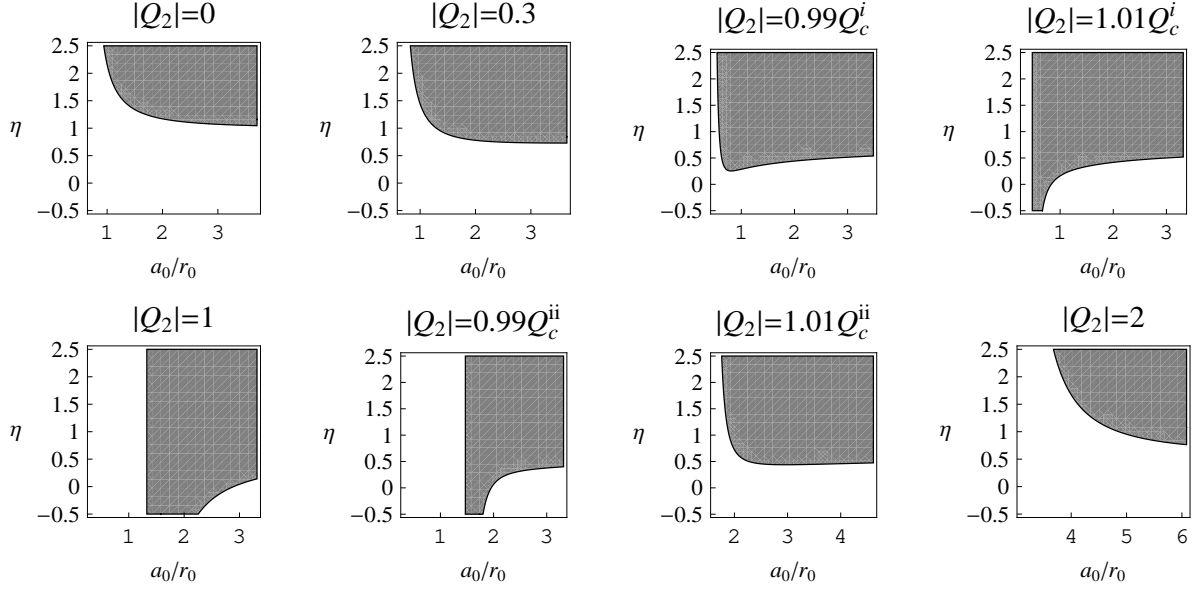}
\caption{Shell around a black hole in Einstein--Maxwell theory. The values of the parameters are $\Lambda r_0^2=-1$, $M_1=0.1$, $Q_1=0$, $M_2=0.5$; the critical values of the charge are $Q_c^{i}=0.4321$ and $Q_c^{ii}=1.4682$.}
\end{figure}
\begin{figure}[t!]
\centering
\includegraphics[width=12cm]{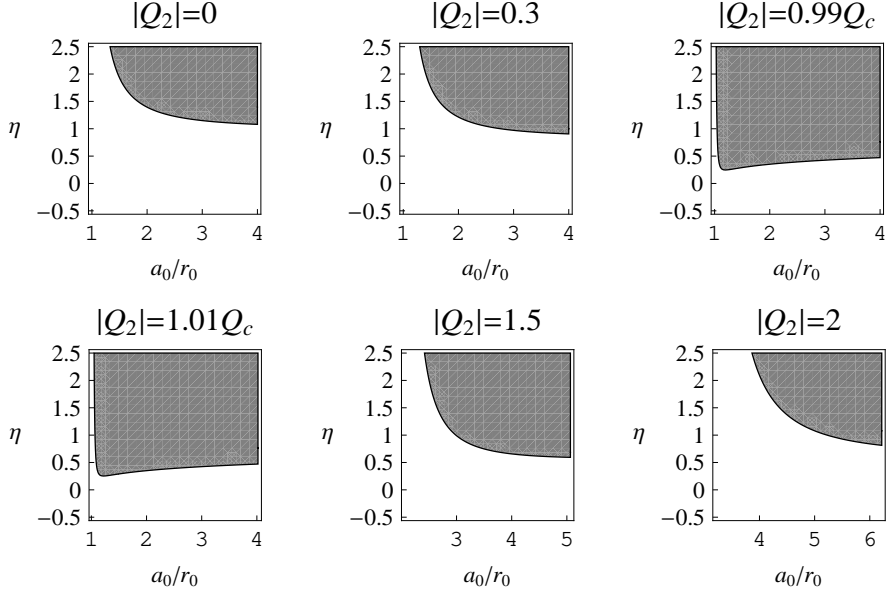}
\caption{Shell around a black hole in Einstein--Maxwell theory. The values of the parameters are $\Lambda r_0^2=-1$, $M_1=0.2$, $Q_1=0$, $M_2=1$; the critical value of the charge is $Q_c=1$.}
\end{figure}
\begin{figure}[t!]
\centering
\includegraphics[width=16cm]{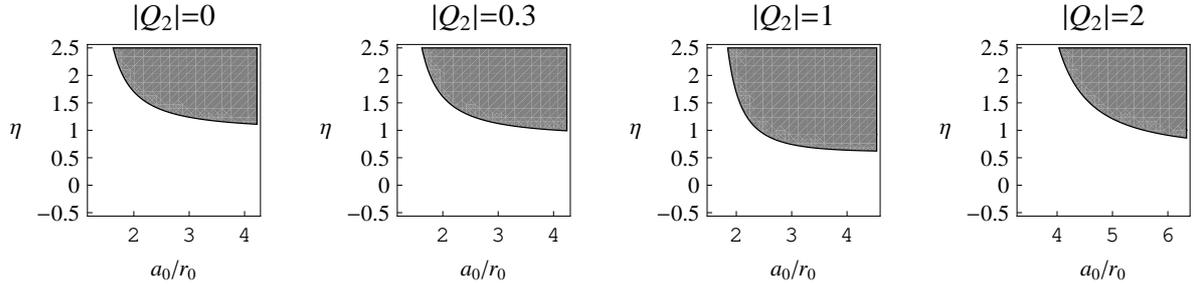}
\caption{Shell around a black hole in Einstein--Maxwell theory. The values of the parameters are $\Lambda r_0^2=-1$, $M_1=0.3$, $Q_1=0$, $M_2=1.5$.}
\end{figure}
\begin{figure}[t!]
\centering
\includegraphics[width=16cm]{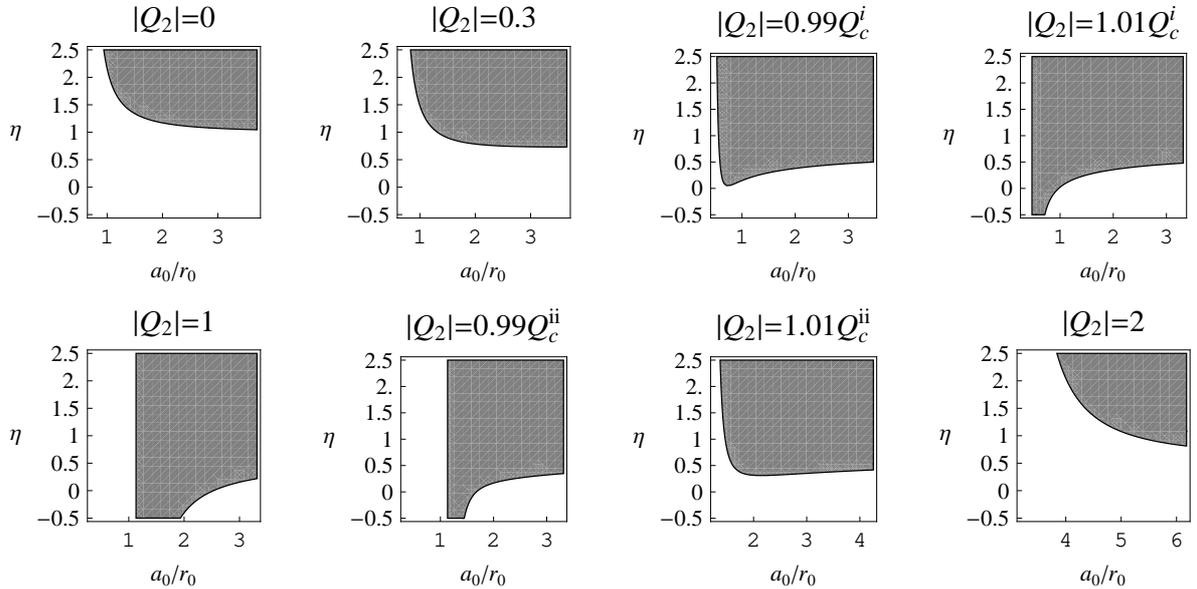}
\caption{Shell around a black hole in Einstein--Born-Infeld theory with $b/r_0=1$. The values of the parameters are $\Lambda r_0^2= -1$, $M_1=0.1$, $Q_1=0$, $M_2=0.5$; the critical values of the charge are $Q_c^{i}=0.4657$ and $Q_c^{ii}=1.2597$.}
\end{figure}
\begin{figure}[t!]
\centering
\includegraphics[width=12cm]{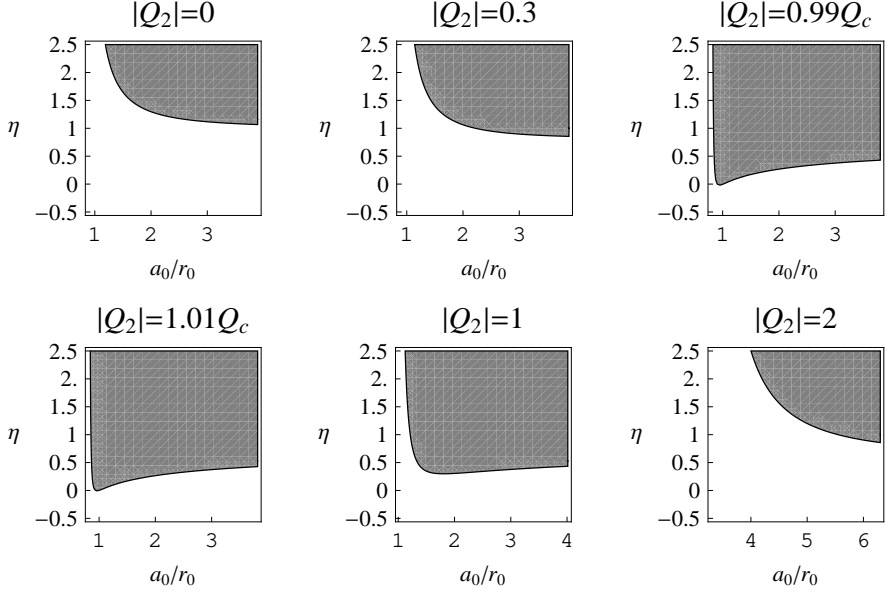}
\caption{Shell around a black hole in Einstein--Born-Infeld theory with $b/r_0=1$. The values of the parameters are $\Lambda r_0^2= -1$, $M_1=0.16$, $Q_1=0$, $M_2=0.8$; the critical value of the charge is $Q_c=0.8944$.}
\end{figure}
\begin{figure}[t!]
\centering
\includegraphics[width=16cm]{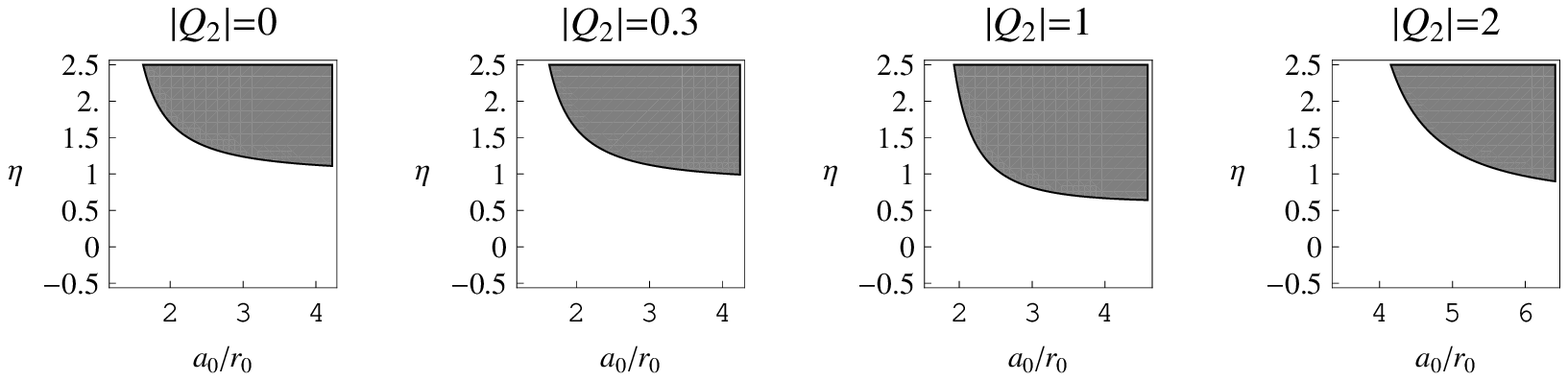}
\caption{Shell around a black hole in Einstein--Born-Infeld theory with $b/r_0=1$. The values of the parameters are $\Lambda r_0^2= -1$, $M_1=0.3$, $Q_1=0$, $M_2=1.5$.}
\end{figure}

In the case of shells around black holes in Einstein--Born--Infeld theory, we fix the scale with the choice of the cosmological constant; the inner metric corresponds to non charged black holes ($Q_1=0$) with different masses, thus the ring charge is simply the charge associated to the exterior metric. We first consider the limit $b\to 0$, which corresponds to Maxwell electrodynamics. The values of the masses and charges of the exterior metrics are shown in Figs. 1 to 3. For low values of the mass $M_2$, two critical values of the charge $Q_2$ exist: $Q_c^{i}$ and $Q_c^{ii}$ are such that for a charge under $Q_c^{i}$ and for a charge beyond $Q_c^{ii}$ there is an event horizon in the original outer metric, while for $0<Q_c^{i}<|Q_2|<Q_c^{ii}$ it presents a naked singularity. For a certain value of the mass, the critical values of the charge fuse into one, and beyond that an event horizon always exists in the original exterior metric for any value of the charge. In the complete manifold, the features of the 
outer original metric, i. e. its horizon structure, determine the behavior (range, location and shape) of the stability regions in parameter space as the charge increases. In particular, only in the case in which the horizon of the original metric is lost, stability could be possible for vanishing or negative $\eta$. In all cases a vanishing charge makes stability compatible only with $\eta>1$. Figs. 1 to 3 illustrate the behavior of the stability regions with an increase of the charge, for a fixed value of the cosmological constant and three different values of the mass. The main difference with the results for the four dimensional case \cite{eisi-bi} is that in three dimensions the evolution of the regions is not monotonous: as the considered charge becomes larger, so that the original outer geometry has an event horizon, the stability regions recover the same form of those corresponding to low values of the charge. 
 
In the case of Einstein gravity coupled to Born--Infeld non linear electrodynamics, the departure from standard Maxwell theory is determined by the parameter $b$, which can be taken non negative without losing generality. The stability study is performed in terms of the dimensionless quantity $b/r_0$.  As before, we consider charged shells around $(2+1)$-dimensional non charged black holes; the results are shown in Figs. 4 to 6. The behavior of the original outer metric with the charge is analogous to that of the Maxwell case; however, for all other parameters fixed, the difference between the values of the two critical charges is smaller in Born--Infeld electrodynamics. More precisely: for a given value of $b/r_0$, the difference between $Q_c^{ii}$ and $Q_c^i$ decreases as the mass $M_2$ is enlarged, while for a given value of $M_2$ this difference becomes smaller as a larger $b/r_0$ is chosen. This is reflected in the evolution of the shape and size of the stability regions as functions of the parameters. 
We display only the results corresponding to $b/r_0=1$. The plots show that the region such that smaller positive and also negative values of $\eta$ are compatible with stability is reduced and finally disappears as the mass $M_2$ is taken larger. Another point to be noted is that, for charges near the critical ones, the stability regions for the same values of $|Q_2|/Q_c$ are in general slightly larger in the case of Born--Infeld electrodynamics, making stability compatible with lower values of the parameter $\eta$.

As a particular case, we can study  bubbles --rings around vacuum-- by taking the inner black hole mass $M_1$ equal to zero.  This can be done for both Maxwell and Born--Infeld theories. Within both theoretical frameworks the results are very similar to those corresponding to rings around black holes, so figures are not included. However, the numerical analysis shows that, in general, for the same values of the ring charge, rings around black holes admit slightly larger stability regions, associated to smaller possible  values of the parameter $\eta$, than rings around vacuum.

\section{Summary}

We have constructed  charged rings around $(2+1)$-dimensional non charged black holes within the framework of Einstein--Born--Infeld theory, and we have studied their stability under linearized radial perturbations. The Einstein--Maxwell case corresponds to the limit in which the Born--Infeld parameter $b$ is zero. We have found that, for fixed $\Lambda r_0^2$ and taking a given $b/r_0$, for low values of the mass $M_2$ of the outer metric, there are two critical values of the charge such that the stability regions are larger for charges between them; these critical charges are no longer present for large values of the mass $M_2$, and the stability regions are smaller in this case. There are ranges of the parameters for which the stability regions include the physically more interesting case $0<\eta<1$, so $\eta $ can be understood as the velocity of sound on the shell. For fixed $M_2$, the critical values of charge get closer to each other as $b/r_0$ grows, reducing  the range of charge corresponding to 
larger stability regions.

We have also considered the particular case of bubbles, that is, charged rings around vacuum. Though not displayed, the numerical results show that, for fixed values of the parameters, the stability regions of rings around vacuum are slightly smaller than those of rings around black holes, in the sense that stiffer matter is needed.

\section*{Acknowledgments}

This work has been supported by Universidad de Buenos Aires and CONICET.

\end{document}